# Spatiotemporal Control of Ultrafast Pulses in Multimode Optical Fibers

Daniel Cruz-Delgado[1]*, J. Enrique Antonio-Lopez[1], Armando Perez-Leija[1], Nicolas K. Fontaine[2], Stephen S. Eikenberry[1], Demetrios N. Christodoulides[3], Miguel A. Bandres[1], Rodrigo Amezcua-Correa[1]

1. CREOL, The College of Optics and Photonics, the University of Central Florida, Orlando, Florida 32816, USA

2. Nokia Bell Labs, 600 Mountain Avenue, Murray Hill, NJ 07974, USA

3. Ming Hsieh Department of Electrical and Computer Engineering, University of Southern California, Los Angeles, California 90089, USA

*Corresponding author. Email: daniel.cruzdelgado@ucf.edu

**Multimode optical fibers represent the ideal platform for transferring multidimensional light states. However, dispersion degrades the correlations between the light's degrees of freedom, thus limiting the effective transport of ultrashort pulses between distant nodes of optical networks. Here, we demonstrate that tailoring the spatiotemporal structure of ultrashort light pulses can overcome the physical limitations imposed by both chromatic and modal dispersion in multimode optical fibers. We synthesize these light states with predefined spatial and chromatic dynamics through applying a sequence of transformations to shape the optical field in all its dimensions. Similar methods can also be used to overcome dispersion processes in other physical settings like acoustics and electron optics. Our results will enable advancements in laser-based technologies, including multimode optical communications, imaging, ultrafast light-matter interactions, and high brightness fiber sources.**

The combination of spatiotemporal light beam synthesis and sophisticated multimode waveguides raises the intriguing possibility of harnessing ultrashort light pulses with rich dynamics in a variety of applications. Yet, an optical pulse that originally exhibits precise spatiotemporal correlations prior to transmission through a medium, invariably loses its initial state due to effects arising from dispersion. Specifically, the spatial and chromatic constituents of light travel at different speeds, leading to undesirable aberrations in the overall pulse structure. Consequently, controlling the energy distribution contained in optical pulses as they traverse multimode channels presents a formidable challenge. This affects diverse technological areas such as ultrafast optics, fiber lasers, communications, and custom light delivery. Addressing this problem through the use of richer pulse light configurations in multimode environments will facilitate investigations of fundamental phenomena such as multi-mode multicolor solitons [1], spatiotemporal probing and coherent control of ultrafast light-matter interactions [2-4]. Recent work on space-time beams incorporates nonseparable spatiotemporal states via generation of localized wavepackets with controllable group velocity [5, 6], spatiotemporal orbital angular momentum (OAM) dynamic beams [7-12, 2], and toroidal vortices [13, 14]. Additionally, major experimental efforts have led to spatiotemporal focusing using pulsed laser sources [15-19], combination of spatial modes in continuous-wave conditions [20], and excitation of supermodes in unbounded planar waveguides [21] and fiber optics [22]. In the context of imaging systems, of special interest are multimode optical fibers, which many consider to be the probe of choice in terms of examining the interior of living organisms and tissues [23, 24], and observing the intricate details of biological systems [25, 26]. To faithfully retrieve the optical information, most methods require prior knowledge of the light scrambling processes by either computing and measuring the transmission matrix [27-29, 22, 20], or by learning the output speckle patterns [30-38], in order to diminish the detrimental effects of random scattering and dispersion. Notwithstanding these advancements, uncontrolled dispersion effects on spatiotemporal pulses in multimode optical fibers continue to limit the ultimate performance of these technologies. Here we use a two-dimensional pulse synthesizer and state-of-the-art spatial-mode multiplexer [11] to mitigate and overcome modal and chromatic dispersion effects. We achieve this by manipulating the spatiotemporal degrees of freedom (DOF) of ultrashort light pulses prior to propagation through multimode optical fibers. Specifically, we can deliver beams with specific spatial profiles and prescribed frequency modulations.

**Experimental Setup**

Figure 1. illustrates this concept and our experimental setup. The first element in our arrangement is a laser source generating a discrete array of evenly spaced pulses with pulse duration of $180 fs$ centered at a wavelength of $\lambda = 1030 nm$ with $6.4 nm$ bandwidth full-width at half-maximum (FWHM), Fig. 1b. We split the pulse train to generate a reference beam and a signal beam. We insert a delay line to vary the reference beam path length, so we can perform holographic interference for examining the output beam. Concurrently, a reconfigurable space-time (ST) module allows us to readily program the spatiotemporal structure of the signal beam

and then precisely couple it into a multimode fiber. The temporal stage of the ST module consists of a two-dimensional (2D) Fourier-transform pulse shaper that controls the spectral content (eigen-frequencies) of the signal pulse and a diffraction grating that decomposes the light pulse into its constituent wavelengths. A cylindrical lens then projects the pulse onto a two-dimensional (2D) spatial light modulator (SLM). Crucially, as different wavelengths impinge upon the horizontal axis of the SLM, the spectral/temporal shaping of the signal beam takes place along the horizontal direction. That is, the horizontal axis of our SLM is calibrated to imprint a particular phase pattern on each wavelength component, while the vertical axis is configured to steer the spectrally modulated pulses towards predefined positions before entering the spatial stage. A dual-lens telescope adjusts the beam size prior to the spatial transformations. The spatial stage consists of a multi-plane light conversion (MPLC) system which adiabatically transforms the spatial profile of the input field into specific mode functions. The MPLC consists of a second SLM and a mirror parallel to each other such that the input beam acquires the phase morphology of six holograms upon bouncing back and forth at six different positions on the SLM. Altogether, this optical configuration allows full control of the temporal and spatial DOF of the pulsed field.

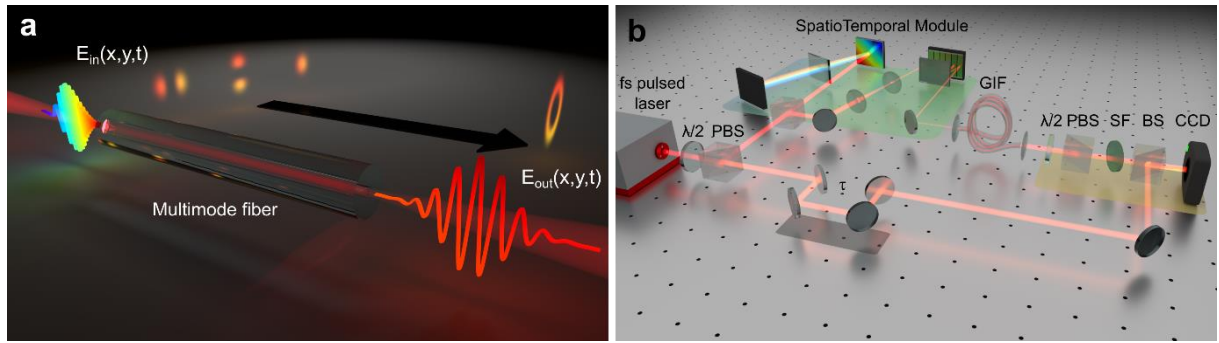

**Fig. 1 | Transport of spatiotemporal light pulses in multimode waveguides**. **a**, A light pulse with a specific spatiotemporal correlation defies modal and chromatic dispersion after propagating in a multimode optical fiber. **b**, A femtosecond pulse is divided into a signal and a reference beam. The signal pulse undergoes a series of transformations in a ST module (green background). Initially a spectral/temporal stage manipulates the signal pulse. At this stage, a 2D Fourier transform pulse shaper tailors the frequency components of the pulse. A multi-plane light conversion system (MPLC) quasi-adiabatically molds the spatial phase and intensity profiles of the wavepacket. The synthesized ST optical field then couples to 2m of graded index fiber. Finally, a frequency resolved tomographic interrogation module (yellow background) analyzes the output light. A CCD records the holographic interference pattern between the reference and signal wave in an off-axis geometry. A time delay on the reference arm scans the time dimension $\tau$.

To demonstrate the capability of our system, we prepare light states with intricate structures and propagate them through a graded index multimode optical fiber. Specifically, we synthesize a single-polarization optical field represented by the spatiotemporal function

$$E_0(x,y,t) = \sum_{m,n,k} F_{m,n,k}(t) e^{i(\Psi_{m,n,k}(t) + \omega_k(t))} HG_{m,n,k}(x,y) e^{i\varphi_{m,n,k}}. \qquad (1)$$

Physically, Eq. (1) describes an entangled (non-separable) optical field formed by a coherent superposition of complex time-varying functions $F_{m,n,k}(t) e^{i\Psi_{m,n,k}(t)}$, that modulate a set of carrier signals of frequency $\omega_k(t)$, each one exhibiting an instantaneous spatially structured Hermite-Gauss profile, $HG_{m,n,k}(x,y)$, and a spatial global phase $\varphi_{m,n,k}$. Note that $(m,n)$ are the indices of the transverse spatial modes. For our experiments, we use a 2m-long graded index optical fiber with a diameter of 11 $\mu m$ and a refractive index contrast $\Delta n = 16 \times 10^{-3}$. This fiber supports six linearly polarized ($LP$) modes at the central wavelength of our source ($\lambda = 1030 nm$). To examine the multidimensional composition of the optical fields, we use a spectrally resolved holographic technique. This allows us to extract the instantaneous complex spatial composition (amplitude and phase) of each spectral component. To achieve this, we record the interference pattern between the signal and reference beams with a charge-coupled device (CCD) camera in an off-axis geometry. By adjusting the time delay of the reference beam, we reconstruct the signal wave dynamics. In addition, we discriminate the spectral components of the spatiotemporal field by controllably tilting a narrow bandpass filter. Furthermore, we access and examine the output polarization light state by employing a tunable polarization filter.

**Results**

To benchmark the performance of our system, we first characterize the response of the fiber by individually exciting each of the first three supported modes $(LP_{01}, LP_{11x}, LP_{11y})$ without altering the spectrotemporal properties of the source pulse, Fig 2. To do so, we shape the spatial profile of the input pulse to match the $HG$ modes ($HG_{00}, HG_{01}, HG_{10}$), which coincide with the $LP$ modes in graded-index fibers. In Figs. 2 (a, c, e) we present the iso-intensity surfaces of the output fields reconstructed for all chromatic components at each specific time and realization. To visualize the impact of chromatic and modal dispersion on the transmitted pulses, we generate the enhanced spectrotemporal maps shown in Figs. 2 (b, d, f). Here, each pixel represents the retrieved instantaneous field at a particular wavelength, expressed in terms of the spatial Laguerre-Gaussian ($LG$) basis, see insets Figs. 2 (b, d, f). Using this $LG$ mode representation, we color-code the angular momentum exhibited by the output fields, $\ell_{-1} = -1$ (yellow), $\ell_0 = 0$ (magenta), and $\ell_1 = 1$ (cyan), where any superposition of these modes is represented by a single color. Typically, in the absence of dispersion, every pulse component (modal and chromatic) travels at the same velocity. However, after propagation through the fiber, modal dispersion causes each mode comprising the pulse to emerge at a different time, producing relative delays along the time axis of the maps. We observe that the spatial mode with vorticity $\ell = 0$ is the fastest, followed by the modes with OAM $\ell = +1$ and $\ell = -1$, respectively. Concurrently, chromatic dispersion produces a positive chirp in the pulses which appears as a positive slope in the spectrotemporal maps.

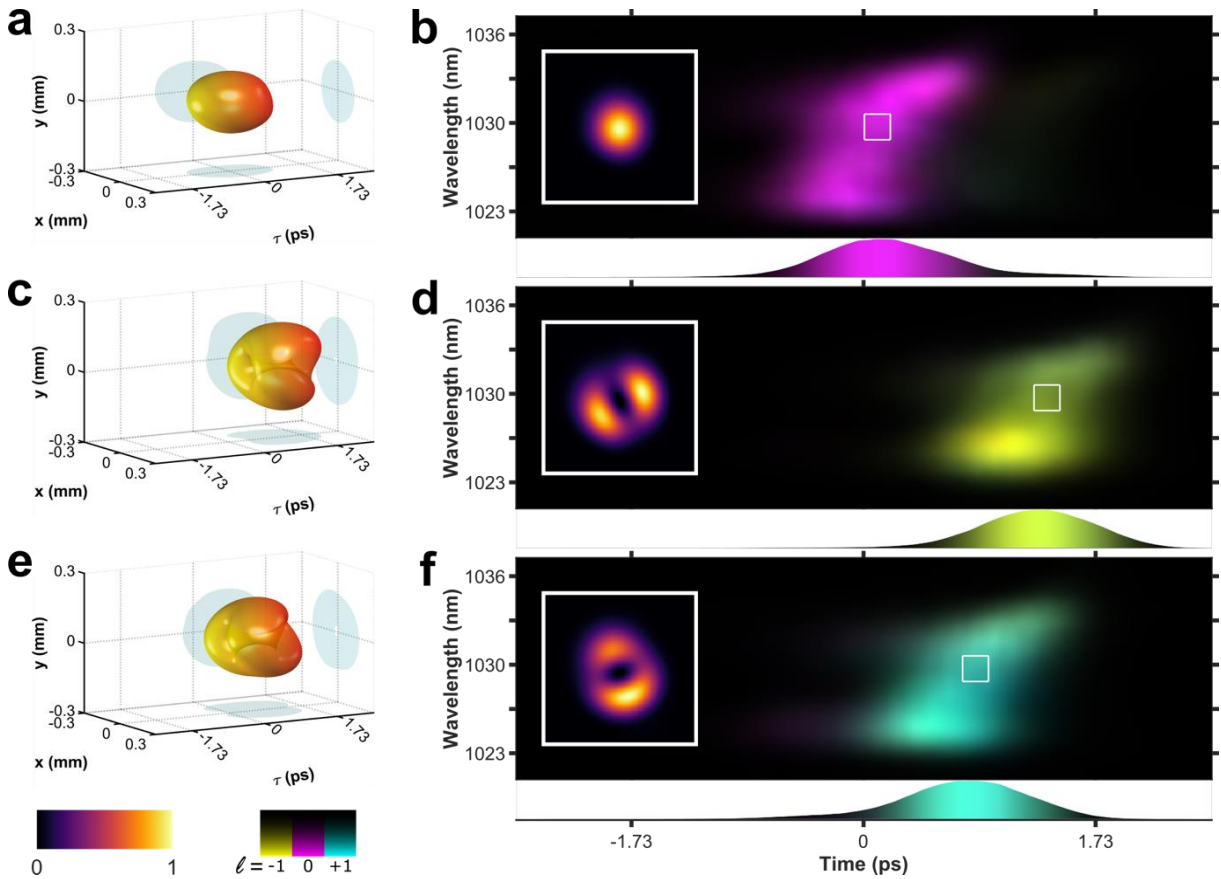

**Fig. 2 | Spatiotemporal response of multimode optical fiber**. **a**, **c**, **e** Measured iso-intensity surfaces of the pulsed field when individually each one of the first three supported modes are excited. **b**, **d**, **f** Color-coded modal decomposition maps after projecting the field onto a set of Laguerre–Gaussian ($LG$) modes. Each mode of this basis is associated with an $CMY$ color ($\ell_{-1} = -1$ (yellow), $\ell_0 = 0$ (magenta), and $\ell_1 = 1$ (cyan)). We observe that the spatial mode with vorticity $\ell = 0$ is the fastest, followed by the modes with OAM $\ell = +1$ and $\ell = -1$, respectively. Chromatic dispersion produces a delay in the spectral components of the pulse, which appears as a positive slope in the shape. In the insets, we present a detailed view of the spatial structure of the propagated pulses.

We now show that we can readily manipulate the spatiotemporal structure of the input pulse to negate the effects of dispersion and to eliminate the relative modal time delays.

To highlight the combined effect of modal and chromatic dispersion on the spatiotemporal pulses, we propagate an optical field composed of all three accessible spatial modes $(LP_{01}, LP_{11x}, LP_{11y})$ that preserve the spectral and temporal properties of the source pulse. In Fig. 3a, we display the measured iso-intensity surface of the propagated field. The spectrotemporal map in Fig. 3b depicts the aberrations in the overall pulse structure. The three modes are delayed relative to one another, while also acquiring positive chirps (represented as positive slopes). In Fig. 3c, we present intensity distributions associated with the selected sections of the measured spectrotemporal map.

Next, we fine-tune the timing between the three input modes $(HG_{00}, HG_{01}, HG_{10})$, in order to counteract the effects of modal dispersion in the fiber. To achieve this, we synthesize a train of three pulses, each assigned a different spatial profile/mode and corresponding propagation constant, with time offsets selected such that they temporally overlap at the fiber output. We delay the $HG_{01}$ and $HG_{10}$ modes respect to the $HG_{00}$ mode by $1.14 ps$ and $0.62 ps$, respectively. Figs. 3d depicts the experimentally measured iso-intensity surface of the output field. In Fig. 3e we display the field's modal composition in the color-coded map. The chromatic superposition of yellow, magenta, and cyan creates a uniform whitish color in the spectrotemporal map (Fig. 3e), indicating that the three modes exit the fiber simultaneously. In the inset of Fig. 3e we present the spatial mode composition of the field retrieved from the white pixel squared region.

In a second experiment, we generate a monolithic structure by sculpting the $HG_{00}$ mode at longer wavelengths of the pulse, while creating an $HG_{01}$ mode at a shorter wavelength, such that the combination produces a single spatiotemporal entity at the fiber output. In other words, we compensate chromatic dispersion by properly manipulating the modal dispersion that the pulse is expected to experience during propagation. The iso-intensity surface corresponding to this case is shown in Fig. 3f. Figure 3g displays the composition of this spatiotemporal pulse, where the magenta and cyan colors represent the two different angular momentum states, OAM value $\ell = 0$ at the longer wavelengths and $\ell = +1$ at the shorter wavelengths, both emerging simultaneously. In the insets of Fig. 3f we show the spatial intensity distribution of the field corresponding to the white square regions.

To further illustrate our system's capabilities, we next simultaneously negate both modal and chromatic dispersion effects. We compensate chromatic dispersion of a pulsed field with negative vorticity ($\ell_{-1} = -1$), and additionally we tailor its spectrum to exhibit a positive cubic frequency function. Figure 4a depicts the iso-intensity surface corresponding to this pulsed vortex field (having a hollow cylindrical shape), while Fig. 4b shows the spectrotemporal configuration. The parabolic-like structure in the spectrotemporal map provides clear evidence of the cubic chromatic function imposed in the field. Figure 4c confirms the vortex attributes of the beam at the selected (white square) region in the spectrotemporal map.

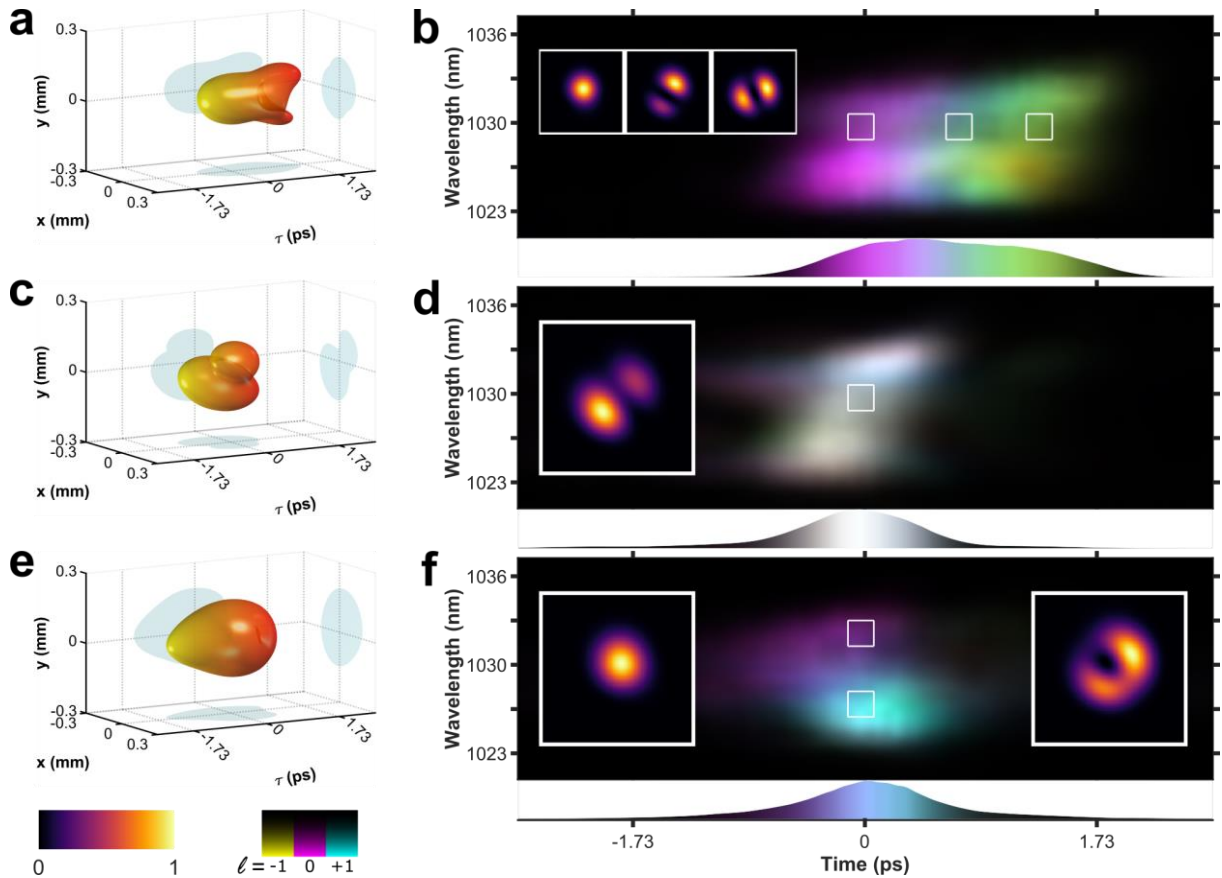

**Fig. 3 | Dispersion mitigation 1**. Measured iso-intensity of **a**, A wavepacket composed of our three accessible spatial modes coupled together. **c**, A field structure made-up of appropriately delayed *LP* modes. **e**, A single spatiotemporal entity comprised of two different spatial modes. Color-coded modal decomposition maps displaying **b**, The three different mode elements delayed and positively chirped. **d**, The three modes arriving simultaneously with positive chirp. **g**, A pulse having two different topological charges, with vorticity value $\ell = 0$ in the longer wavelengths and $\ell = +1$ in the shorter wavelengths. In the insets we present the spatial intensity structure of the wavepackets.

Finally, we demonstrate the capabilities of our system by fully mitigating dispersion. To accomplish this, we tailor a temporal pulse train in which we swap the characteristic arrival time of the guided modes while individually tailoring the spectral components of each element. In Figure 4d we present the corresponding iso-intensity surfaces displaying the spatial patterns of the LP modes in the predefined temporal order. We tailor the first element to exhibit positive vorticity ($\ell_1 = 1$) and a negative cubic chromatic function (parabolic-like shape pointing to the left of the map). The second pulse has negative vorticity ($\ell_{-1} = -1$) and a constant phase.

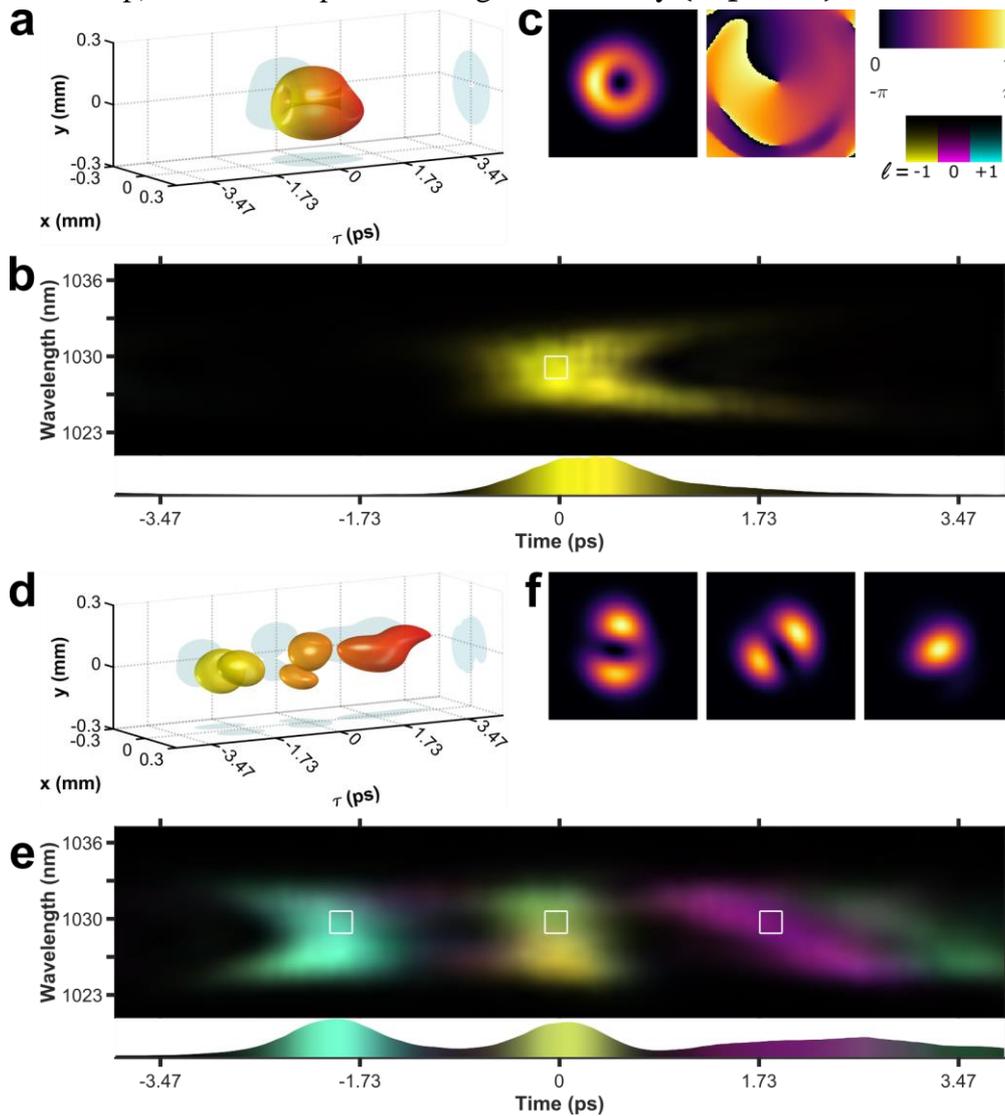

**Fig. 4 | Dispersion mitigation 2**. Measured iso-intensity of the wavepackets. **a**, A pulsed field with spatial negative vorticity ($\ell = -1$). **d**, Train of three pulses with a different spatial mode structure matching the $LP$ modes guided in fiber. Here we set the temporal order of the modes ad libitum. Color-coded modal decomposition maps displaying **b**, The OAM mode $\ell = -1$. Additionally, the cubic spectral function is manifested as a parabola-like shape. (**e**) A precise spectral function fashioned on each element of the train of pulses, the $LP_{11x}$ mode with a cubic spectral function, the $LP_{11y}$ mode with no frequency chirp, and the $LP_{01}$ mode with negative chirped. **c**, **f**, Spatial profile intensity of selected regions of the spatiotemporal structures.

The third element has vorticity ($\ell_0 = 0$) and negative quadratic chromatic function (negative slope). Figure 4e shows the imposed frequency functions for each component. The spatial profiles in Fig. 4f demonstrate the transition of the LP mode content across different spectral/temporal regions. These results clearly demonstrate that our platform can establish a new design paradigm for ultrafast fiber applications and fiber amplifiers.

In summary, we experimentally demonstrated the ability to deliver a rich variety of spatiotemporal states of pulsed light in order to overcome the adverse effects encountered in a complex environment like a multimode waveguide. Our methodology allows one to synthesize arbitrary multidimensional light pulses in a robust and versatile manner. This has implications for a broad range of applications such as fiber-based multimode imaging systems, long-distance communications, ultrafast light-matter interactions, optical fiber amplifiers, and multidimensional information encoding.

**Methods**

We employ a pulse laser source centered at $\sim 1030\ nm$, $180 fs$ ($\sim 6.4\ nm$ FWHM, $sech^2$-shaped pulse) and repetition rate of $40\ MHz$ (Origami NKTPhotonics). We use a half-wave plate (AHWP05M-980 Thorlabs) and a polarization beam splitter (PBS123 Thorlabs) to controllably divide the beam into reference and signal pulses. We temporally modulate the signal pulse train with a two-dimensional Fourier transform pulse shaper in a folded configuration. We use a ruled reflective diffraction grating ($1200\ grooves/mm$, Blaze wavelength $1\ \mu m$ Thorlabs). A C-coated cylindrical lens with $30\ cm$ focal length (LJ1558RM-B Thorlabs). Additionally, we employ a $1920 \times 1080$ pixels reflective SLM, pixel size $8\ \mu m$ (PLUTO-2.1-NIR Holoeye). We direct the modulated pulse coming out of the two-dimensional Fourier transform pulse shaper to the next stage using a 50/50 beam splitter (BS014 Thorlabs). We adjust the size of the field with a two-lens system composed of two C-coated plano-convex lenses $f_1 = 3.5\ cm$ (LA1027-C Thorlabs) and $f_2 = 30\ cm$ (LA1484-C Thorlabs). At the last stage, we spatially engineer the field profile with a MPLC system using a $1920 \times 1080$ pixels reflective SLM, pixel size $8\ \mu m$ (PLUTO-2.1-NIR Holoeye) and a squared mirror (PFSQ05-03-P01 Thorlabs).

For our experiments, we use 2m-long of graded index optical fiber home drawn with a diameter size of $11\ \mu m$ and a refractive index contrast $\Delta n = 16 \times 10^{-3}$. We wind our fiber on a $20\ cm$ diameter coil. We use a B-coated aspheric lens with $4.6\ mm$ focal length (CFC5A-B Thorlabs) to couple the synthesized fields in our graded index fiber. Finally, we use a B-coated aspheric lens with $11\ mm$ focal length (C397TMD-B Thorlabs) to uncouple the light.

To analyze the spatiotemporal structure of the fields, we use the reference pulse with controllable time delay. We use a second set of half-wave plate and a polarization beam splitter to adjust the reference pulse's intensity. We resolve the frequency components of the signal pulse by employing a spectral filter centered at $1047.1 nm$, $4\ nm$ FWHM bandwidth (LL01-1047-12.5 Semrock) mounted on a rotation stage (Thorlabs ELL14). We recombined the reference and signal arms using a 50/50 beam splitter (BS005 Thorlabs). We distinguish the polarization state using a system composed of a half wave plate (AHWP05M-980 Thorlabs) and a polarization beam splitter (PBS053 Thorlabs). Finally, we record the reference and signal interference pattern on a CCD camera $2048 \times 1088$ pixels, $5.5\ \mu m$ pixel size (Beamage-3.0 Gentec). In order to capture the dynamics of the field, a linear translation stage driven by an actuator (Z825B Thorlabs). We scan the time dimension by introducing a time delay $\tau$ on the reference arm at time steps of ~66.66 $ps$ (20 $\mu m$ optical path length steps), while we interrogate the frequency components by rotating the spectral filter at angle steps of 0.5°.

**Data availability**

The data that support the findings of this study are available from the corresponding authors upon reasonable request.

**Code availability**

All the relevant computing codes used in this study are available from the corresponding author upon reasonable request.

**Acknowledgements**

This effort was sponsored, in part, by the Army Research Office of Scientific Research (W911NF1710553 and W911NF1910426); NASA (80NSSC21K0624); Department of the Navy, Office of Naval Research, (N00014-20-1-2789); the National Science Foundation (EECS-1711230); the Simons Foundation (733682); the US-Israel Binational Science Foundation (BSF; 2016381).

**Author contributions**

All authors contributed to all aspects of this work. D.C.D. performed the experiments in consultation with all the team members.

**Competing interests**

Authors declare that they have no competing interests.

**Material & Correspondence**

Correspondence and requests for materials should be addressed to Daniel Cruz-Delgado.